\documentclass[aps,showpacs,preprintnumbers,amsmath,amssymb,nofootinbib,superscriptaddress]{revtex4}
\usepackage{hyperref}
\usepackage{float}
\usepackage{graphics,epsfig}
\usepackage{bm}
\usepackage{slashed}
\usepackage{graphicx}%Include figure files
\usepackage{color}%
\newcommand{\abs}[1]{\lvert#1\rvert}

\newcommand{\no}{\nonumber}
\usepackage{dcolumn}% Align table columns on decimal point
\providecommand{\U}[1]{\protect\rule{.1in}{.1in}}

\begin{document}
\title{Building a Holographic Superconductor
with Higher-derivative Couplings}

\author{Xiao-Mei Kuang}
\email{kuangxiaomei@sjtu.edu.cn}
\affiliation{INPAC, Department of Physics and Shanghai Key Lab
for Particle Physics and Cosmology, Shanghai Jiao Tong University,
Shanghai 200240, China.}
\author{Eleftherios Papantonopoulos}
\email{lpapa@central.ntua.gr}
\affiliation{Department of Physics, National Technical University of
Athens, GR-15780 Athens, Greece.}
\affiliation{INPAC, Department of Physics and Shanghai Key Lab
for Particle Physics and Cosmology, Shanghai Jiao Tong University,
Shanghai 200240, China.}
\author{George Siopsis}
\email{siopsis@tennessee.edu}
\affiliation{Department of Physics and
Astronomy, The University of Tennessee, Knoxville, TN 37996 - 1200, USA.}
\author{Bin Wang}
\email{wang_b@sjtu.edu.cn}
\affiliation{INPAC, Department of Physics and Shanghai Key Lab
for Particle Physics and Cosmology, Shanghai Jiao Tong University,
Shanghai 200240, China.}
%\author{Xiao-Mei Kuang$^1$, Eleftherios Papantonopoulos$^{1,2}$, George Siopsis$^3$, Bin Wang$^1$}
%\affiliation{$^1$INPAC, Department of Physics and Shanghai Key Lab
%for Particle Physics and Cosmology, Shanghai Jiao Tong University,
%Shanghai 200240, China\\
%$^2$Department of Physics, National Technical University of
%Athens, GR-15780 Athens, Greece\\$^3$Department of Physics and
%Astronomy, The University of Tennessee, Knoxville, TN 37996 -
%1200, USA}

\date{October 2013}
%\vspace*{0.2cm}
\begin{abstract}
%\baselineskip=0.6 cm
%\begin{center}
%{\bf Abstract}
%\end{center}
We discuss the gravitational dual of a holographic superconductor
consisting of a  $U(1)$ gauge field, a complex scalar field
coupled to a charged AdS black hole and a higher-derivative coupling between the
$U(1)$ gauge field and the scalar with coupling constant $\eta$. In the presence of a magnetic field, the
system possesses localized spatially dependent droplet solutions
which,  in the low temperature limit, have smaller critical
temperature for $\eta >0$ than the droplet solutions without the interaction
term ($\eta =0$).
% reminiscent to the formation of FFLO states  in the generalized Landau-Ginzburg Theory.
In the weak magnetic field
limit, the opposite behavior is observed: the critical temperature increases as we increase $\eta$. We also calculate the energy gap in the probe limit and find that it is larger for $\eta <0$ than the energy gap in the conventional case ($\eta =0$).
\end{abstract}

\pacs{11.25.Tq, 04.70.Bw, 74.20.-z}\maketitle
%\newpage
%\vspace*{0.2cm}

\section{Introduction}

%%%%%%%%%%%%%%%%%%%%%%%Indean%%%%%%%%%%%%%%%%%%%%%%%%%%%

The gauge/gravity duality is a powerful method of studying strongly
coupled phenomena using dual weakly coupled gravitational systems
 \cite{Maldacena:1997re}. This duality which can be
considered as one of the most successful applications of string
theory. One of the interesting applications is on  condensed  matter systems
(for a review, see \cite{Hartnoll:2009sz}). In trying to understand
condensed matter phenomena, e.g., superconductivity, using the
gauge/gravity duality, two main approaches have been followed. One is a more phenomenological approach (termed
\emph{bottom-up} approach), in which terms in the action are postulated and their effects studied, postponing the derivation of such terms and their relative strength from a full quantum theory to the future.
In the other approach (termed \emph{top-down} approach), one starts with the full quantum theory (string/M theory) and uses it to derive all terms in the action. Although the latter is a more rigorous approach, the range of applications is limited. In our view, progress needs to be made in both directions.

In the bottom-up approach, the simplest holographic superconductor
model is described by an
Einstein-Maxwell-scalar field theory with a negative cosmological
constant \cite{Hartnoll:2008vx,Hartnoll:2008kx}. The gravity dual of the holographic superconductor is
an Abelian-Higgs model with a stationary black hole metric in which the
scalar field condenses below a certain critical temperature.
The mechanism of instability has a geometric origin
\cite{Gubser:2005ih,Gubser:2008px}. If the charge of the scalar
field is large enough, then the effective mass of the scalar field
$m^2_{\text{eff}}=m^2+q^2g^{tt}A^2_t$, where $A_t$ is the electrostatic
potential, can drop below the Breitenlohner-Freedman bound near the horizon of the black hole signaling the breaking of an Abelian
gauge symmetry outside the black hole horizon. Since the space is
asymptotically  AdS, the scalar field is trapped outside the
black hole horizon resulting in the formation of a  condensate which
destabilizes the configuration, and a hairy black hole forms
below the critical temperature (for a review, see \cite{Horowitz:2010gk},  and
references therein).

This phenomenological approach has the virtue of simplicity, but
does not capture all the underlying features of the gauge/gravity
duality including quantum effects. This is hard to implement, as it remains a challenge to embed this model in a quantum system (string/M-theory).
In the top-down approach, one aims at finding
 exact solutions of $D=11$ and type IIB
supergravity. The latter approach is mainly based on Kaluza-Klein
truncations of supergravity theories. Fully back reacted
solutions of $D=11$ supergravity  describing
holographic superconductors in three spacetime dimensions have been found
\cite{Gauntlett:2009zw,Gauntlett:2009dn}. These
models, which are consistent truncations of string/M theories reduced to four spacetime dimensions, contain a large
number of scalar, gauge fields and high derivatives of them, which
need to be constrained in order to make the models tractable
\cite{Gauntlett:2009bh}.

Interesting physics arises when an external
magnetic field is applied to a superconductor. As
the temperature is lowered and the external
magnetic field becomes stronger, the
superconductors expel the magnetic field. This is
the well-known Meissner effect.
 Depending on their behavior in the presence of an external magnetic field, ordinary
  superconductors are classified into two categories, namely type I and type II.
   In type I superconductors, for fields stronger than a critical value ($ B>B_c $), a first-order transition
occurs   from the superconducting phase to the normal phase. On the other hand, in type II superconductors,
   a gradual second-order phase transition occurs, and the material
   ceases to be a superconductor for $ B>B_{c2} $, where $ B_{c2} $ is the upper critical field strength.

Holographic superconductors in the presence of external magnetic
field have been discussed in the bottom-up approach
\cite{Albash:2009ix,Ge:2010aa,Domenech:2010nf,Montull:2012fy}.
It was found that for a non-zero external
magnetic field in two spatial dimensions, it
is inconsistent to have non-trivial spatially
independent solutions on the boundary. Two classes of localized solutions were found, the droplet
\cite{Albash:2008eh}, and vortex solutions with
integer winding number
\cite{Albash:2009iq,Montull:2009fe,
Maeda:2009vf}.

In the top-down approach, the effects of an external magnetic field
were discussed in \cite{Ammon:2008fc,Ammon:2009fe}. A model was proposed \cite{Ammon:2008fc} which    had an explicit field theory
realization as strongly coupled  $\mathcal{N}=2$ super Yang-Mills
theory  with flavor.   Using gauge/gravity duality, with the probe
of two flavor D7--branes in the AdS black hole background, it was
shown that the system  underwent a second-order phase transition
to a new state which  was a p-wave superconductor because of
the presence of flavor symmetry. Subsequently in
\cite{Ammon:2009fe}, the Meissner effect was studied by introducing an
external magnetic field.

In all the above approaches, the effects of an
external magnetic field on the holographic
superconductor were considered within the standard
Maxwell electrodynamics and coupling to charges. There are various
reasons to consider higher-derivative terms as
corrections to the usual Maxwell field couplings.
First of all these couplings appear in a
consistent  truncation of string/M theories.
Secondly,  using the usual Maxwell theory it is
widely believed that holographic superconductors
are of type II rather than type I. It would be
interesting to see if these terms affect  the
behavior of holographic superconductor under the
influence of an external magnetic field. In this
direction \cite{Roychowdhury:2012hp} (see also
\cite{Roychowdhury:2012vj}), by introducing  higher-derivative corrections to the usual Maxwell
action, it was found that the value of the
critical
 field strength $ B_c $ was affected, indicating  that the presence
 of these terms may alter the nature of the phase transition from
 the superconducting to the normal state. Further work is needed to
 understand whether these terms can give a first-order transition.

Our main motivation to consider higher-order
terms comes from the recent progress in
high-$T_c$ superconductors. It was found that,
when a strong external magnetic field coupled to
the spins of the conduction electrons is applied
to a superconductor, inhomogeneous phases
appear as the temperature is lowered. This
results in a separation of the Fermi surfaces
corresponding to electrons with opposite spins
(for a review see \cite{Casalbuoni:2003wh}), and
the electron pairing is destroyed, resulting in a first-order transition from the
superconducting state to the normal state. As has been
shown by Fulde and Ferrell \cite{Fulde},
and Larkin and Ovchinnikov \cite{Larkin}, a
new state (the FFLO state) can form with a
modulated order parameter in a weakly coupled BCS system.

A way to understand the formation of  these inhomogeneous states in a strongly coupled system is to generalize the potential of holographic superconductors by including terms containing higher-derivative couplings of the scalar field to the gauge potential. Such terms should arise from quantum corrections in a top-down approach.
In the bottom-up approach followed here, we are led to consider the effective potential
\begin{equation}
 V(\Psi)= m^2 |\Psi^2| + \eta \abs{F^{\mu\nu}D_\nu\Psi}^2+\eta' \abs{F_{\mu\nu}D^\mu
 (F^{\nu\kappa}D_{\kappa}\Psi)}^2 + \dots~,\label{holpot}
\end{equation}
where $F=dA$ is the field strength of a U(1) gauge field $A_\mu$, $\Psi$
is a charged complex scalar field of mass $m$ and charge $q$, and
$D_\mu =\nabla_\mu-iqA_\mu$.
In this paper, we limit ourselves to the case in which all couplings, except $\eta$, are set to zero  ($\eta'
= \dots = 0$), leaving a more complete discussion to future work.
We study the effects of allowing the gauge field
to develop both an electric and a magnetic component. In the absence of the higher-order interaction
($\eta=0$), we reproduce the droplet solution discussed in
\cite{Albash:2009iq}. In the case of non-vanishing coupling constant $\eta$,
we find that, as the magnetic field becomes stronger, the
droplet solution becomes more inhomogeneous. Moreover, as the temperature
is lowered, there is a critical value $ B_c $ of the magnetic field above which the
critical temperature is higher than in a system with $\eta=0$.

In the case of weak magnetic field, and with
$\eta=0$, we reproduce the results of Ref.\ \cite{Hartnoll:2008vx}. If we
switch on the higher-derivative coupling, we find that the system
undergoes a phase transition at a higher critical temperature.
This gives an interesting and unexpected result: as the strength
of the higher-derivative interaction increases, the gap becomes
smaller. We find a gap $E_g \sim 0.6 T_c$ compared to $E_g \sim
0.8 T_c$ when $\eta=0$. Recently, there has been a renewed
interest in this problem. In \cite{Horowitz:2013jaa}, a periodic
potential was introduced, and a value of $E_g \sim 0.4 T_c$ was found, which is closer to the experimental value.
It would be interesting to see if we can attain this value in our model as well, by tuning the coupling constant $\eta$.

The paper is organized as follows. In section \ref{sec:2}, we
present the details of the model. In section \ref{sec:3}, we
consider droplet solutions, and calculate the critical temperature
with the electromagnetic field having both electric and magnetic
components. In section \ref{sec:4}, we go below the critical
temperature assuming that the magnetic field is weak and calculate
the energy gap in the probe limit. Finally in section \ref{sec:5},
we discuss our conclusions.

\section{The Model}
\label{sec:2}

We consider the action
\begin{equation}\label{action}
 S=\int d^4x \sqrt{-g}\left[\frac{R+6/L^2}{16\pi
 G}-\frac{1}{4}F_{\mu\nu} F^{\mu\nu}
 -\abs{D_\mu\Psi}^2-V(\Psi)\right]~,
\end{equation}
where the potential term is given by (\ref{holpot}) with
 $\eta' = \dots =0$.
 The field equations are:
\begin{itemize}
\item
the Einstein equations
\begin{equation}
R_{\mu\nu}-\frac{1}{2}Rg_{\mu\nu}-\frac{3}{L^2}g_{\mu\nu}=8\pi G
T_{\mu\nu}~, \label{einequation}
\end{equation}
\item
the Maxwell equations
\begin{align}\label{Maxw}
 \nabla_{\mu}F^{\mu\nu}&+\frac{\eta}{\sqrt{-g}}\partial_\mu\left[\sqrt{-g}(D_\kappa \Psi)(D_\lambda
 \Psi)^*\left(g^{\kappa\nu}F^{\mu\lambda}-g^{\kappa\mu}F^{\nu\lambda}+g^{\nu\lambda}F^{\mu\kappa}-
 g^{\mu\lambda}F^{\nu\kappa}\right)\right]\nonumber\\&=iq\left[\Psi^* (D^\nu
 \Psi)-\Psi(D^\nu
 \Psi)^*\right]+iq\eta g_{\mu\rho}F^{\rho\nu}\left[F^{\mu\kappa}\Psi^*
 (D_\kappa\Psi)-F^{\mu\lambda}\Psi
 (D_\lambda\Psi)^*\right]~,
\end{align}
\item and the scalar field equation
\begin{align}\label{eqPsi}
-\frac{1}{\sqrt{-g}}\partial_\mu
\bigl[&\sqrt{-g}g^{\mu\nu}\bigl(\partial_\nu\Psi-iqA_\nu
\Psi\bigr)\bigr]+iqg^{\mu\nu}A_\nu\left(\partial_\mu\Psi-iqA_\mu\Psi\right)+m^2\Psi\no\\&-\frac{\eta}{\sqrt{-g}}\partial_\mu
\left[\sqrt{-g}
g_{\kappa\lambda}F^{\kappa\nu}F^{\lambda\mu}\left(\partial_\nu\Psi-iqA_\nu
\Psi\right)\right]+iq\eta g_{\kappa\lambda}F^{\kappa\nu}F^{\lambda\mu}A_\mu
\left(\partial_\nu\Psi-iqA_\nu \Psi\right)=0~.
\end{align}
\end{itemize}
%where $D^\nu=\nabla^{\nu}-iqA^{\nu}$.
In this work, we will set $L=1$, $8\pi G=1$, $q=1$.

We note that the presence of the coupling constant $\eta$ adds new
terms in the field equations which makes the system of the
differential equations highly non-trivial.
%This gives us the justification in choosing the $g_2=0$.
However, in spite of the
complexity of the system, we were able to perform numerical calculations and also calculate the critical
temperature analytically in the weak field limit.

\section{The critical temperature}
\label{sec:3}

%In the previous section we showed that for spatially independent solutions in the presence only of an electric field the critical
% temperature is enhanced with the presence of the interaction term.
%The original motivation of introducing a potential term  in the Landau-Ginzgurg theory was to study
%the behavior of a superconductor in the presence of a strong
%magnetic field and the generation of anisotropic states in the low
%temperature limit.
In this section, we switch on a magnetic field
and study the effect of spatially dependent solutions on the
critical temperature.

Spatially dependent solutions in the presence of a magnetic
field were  studied in
\cite{Albash:2008eh,Albash:2009ix,Albash:2009iq}. Solving the
Maxwell-scalar equations in a dyonic black hole background
localized droplet and vortex solutions were found and  their
$(B,T)$ phase diagram was studied. In \cite{Hartnoll:2008kx},
with the addition of an external magnetic field, it was argued that the
holographic superconductor is of type II as the magnetic field is
lowered. An analytic study of a holographic superconductor was
carried out in \cite{Ge:2010aa}, where the upper critical magnetic
field was calculated analytically. Vortex lattice solutions in the
presence of a magnetic field were presented in
\cite{Maeda:2009vf}.

To study the effect of the interaction term (with
 $\eta \neq 0$), we will consider  the backreaction
of the Maxwell field to gravity. The Einstein-Maxwell field
equations (\ref{einequation}) and (\ref{Maxw}) at the critical
temperature are solved by a dyonic black hole (i.e., one with
both an electric and a magnetic charge). The metric reads
\cite{Romans:1991nq}
\begin{equation}\label{dynoic}
ds^2= \frac{dr^2}{r^2h(r)} + r^2 \left[ - h(r) dt^2 + dx^2
+dy^2\right] \ \ , \ \ \ \ h(r)=1 -
\frac{r_+^3}{r^3}+\frac{(\lambda^2+\mathcal{B}^2)r_+^4}{4r^3} \left( \frac{1}{r}-\frac{1}{r_+} \right)~.
\end{equation}
%where we have set $L=1$.
The gauge field is given by
\begin{equation}\label{gaugefield}
A_\mu=A_t(r)dt+A_y(x)dy=\lambda r_+\left( 1-\frac{r_+}{r} \right) dt+\mathcal{B}r_+^2xdy~,
\end{equation}
where according to the AdS/CFT dictionary, the charge density and the chemical potential on the boundary are given by $\rho = \lambda r_+^2$ and $\mu = \lambda {r_+}$, respectively. $\mathcal{B}r_+^2$ is the magnitude of the external magnetic field (in the radial direction, $r$, perpendicular to the boundary $xy$-plane).
At the horizon, $r=r_+$, we have $A_t =0$.
The Hawking temperature of the black hole is
\begin{equation}
T=\frac{3r_+}{4\pi}\left( 1-\frac{\lambda^2+\mathcal{B}^2}{12} \right)~.
\end{equation}
In the above background with a magnetic field
$\mathcal{B}$, we will investigate the spatially dependent
solutions of the field equations by considering the ansatz
$\Psi=\Psi(r,x)$.
Near the
AdS boundary, the asymptotic behavior is
$\Psi=\frac{\Psi_1}{r^{\Delta_-}}+\frac{\Psi_2}{r^{\Delta_+}}$
with $\Delta_\pm=\frac{3}{2}\pm\frac{1}{2}\sqrt{9+4 m^2}$.
Both $\Psi_1$ and $\Psi_2$ correspond to normalizable modes, therefore we obtain two different systems (labeled by $\Delta_\pm$) in which one mode is
the source and the other gives the vacuum expectation value of the
dual operator of dimension $\Delta_i$ ($\Psi_i\sim \langle\mathcal{O}_i\rangle$).

Right below the critical temperature, we may regard the scalar
field as  a perturbation in the gravitational background of
(\ref{dynoic}) and (\ref{gaugefield}). Then the scalar field equation in the coordinate $z=\frac{r_+}{r}$
becomes
 \begin{align}\label{scalareq}
&\left[1-\frac{\eta}{r_+^2} z^4
 A_t'^2 \right] \Psi''+\left[\left(\frac{h'}{h}-\frac{2}{z}\right)- \frac{\eta}{r_+^2} z^4\left(\frac{h'}{h}+\frac{2}{z}\right)
 A_t'^2 -\frac{2\eta}{r_+^2} z^4
 A_{t}'A_{t}'' \right]\Psi'\no\\
 &+\frac{1+\eta \mathcal{B}^2 z^4}{h}\partial_{x}^{2}\Psi+\left[-\frac{m^2}{z^2}+\frac{1-\frac{\eta}{r_+^2} z^4 A_{t}'^{2} }{r_+^2 h}A_{t}^{2} -\mathcal{B}^2
 x^2
 -\eta\mathcal{B}^{4} z^{4}x^{2}\right]\frac{1}{h}\Psi=0~,
 \end{align}
where
%$h(z)=1+\frac{\lambda^2+\mathcal{B}^2}{4}z^4-(1+\frac{\lambda^2+\mathcal{B}^2}{4})z^3$ and
the prime denotes differentiation with respect to $z$. Without
loss of generality, we can take $\Psi$ to be real and separate
$\Psi$ into functions of a single variable,
 \begin{equation}\label{sePsi}
 \Psi(z,x)=\Psi_1 (z) \Psi_2 (x)~.
 \end{equation}
Combining equations (\ref{scalareq}) and (\ref{sePsi}), we obtain
\begin{align}\label{Psi}
h&\left[1-\frac{\eta}{r_+^2} z^4
 A_{t}'^{2}\right] \Psi_{1}''+\left[\left(\frac{h'}{h}-\frac{2}{z}\right)- \frac{\eta}{r_+^2} z^4 \left(\frac{h'}{h}+\frac{2}{z}\right)
 A_{t}'^{2}-\frac{2\eta}{r_+^2} z^4
 A_{t}'A_{t}''\right]h\Psi_{1}'\no\\
 &+\left[-\frac{m^2}{z^2}+\frac{ 1-\frac{\eta}{r_+^2} z^4 A_{t}'^{2}}{ r_+^2 h} A_{t}^{2} \right]\Psi_{1}
 +\frac{\Psi_{1}(1+\eta \mathcal{B}^2 z^4)}{\Psi_{2}}\left[\Psi_2''-\mathcal{B}^2
 x^2\Psi_{2}\right]=0~.
  \end{align}
%where the dot denotes differentiation with respect to $x$. The
For consistency  of this equation, we need
\begin{equation}\label{Psi2a}
 \Psi_2''-\mathcal{B}^2 x^2\Psi_2=-k^2\Psi_2~,
 \end{equation}
where $k^2$ is an arbitrary parameter. This equation is of the form of Schr\"odinger's equation for a simple harmonic oscillator.
The eigenvalues are
\begin{equation}\label{Psi2b}
k^2 = (2n+1) \mathcal{B} \ \ , \ \ \ \ n =0,1,2,\dots ~,
\end{equation}
and the corresponding eigenfunctions are given in terms of Hermite polynomials, $H_n (\sqrt{2\mathcal{B}} x)$.
Note  that the $x$-dependent part of the scalar field
obeys the same equation as the one found  in \cite{Albash:2008eh}.
As explained in \cite{Albash:2008eh}, we ought to choose the ground state ($n=0$) to find the critical temperature.

Substituting the $x$-dependent equation (\ref{Psi2a}) into
(\ref{Psi}), we obtain  the field equation for $\Psi_1 (z)$ which
reads
 \begin{eqnarray}\label{eqPsi1}
\left[1-\eta \lambda^2 z^4
\right] \Psi_{1}''+\left[\left(\frac{h'}{h}-\frac{2}{z}\right)-\eta\lambda^2 z^4
 \left(\frac{h'}{h}+\frac{2}{z}\right)\right]\Psi_{1}' & & \no\\
 +\left[\frac{2}{z^2h}+\frac{\lambda^2(1-\eta\lambda^2 z^4)(1-z)^2}{h^2}-\frac{\mathcal{B}(1+\eta
 \mathcal{B}^2 z^4)}{h}\right]\Psi_{1}&=&0~,
  \end{eqnarray}
where we have chosen  $m^2 = -2$ for definiteness, and expressed $A_t $
using  (\ref{gaugefield}).

Observe that new $\eta-$dependent terms appear in the field
equation for the scalar field compared to the scalar equation in
\cite{Albash:2008eh}. They give a different $z-$dependent
behavior of the scalar field, as we will discuss below.

\begin{figure}
\center{
\includegraphics[scale=0.55]{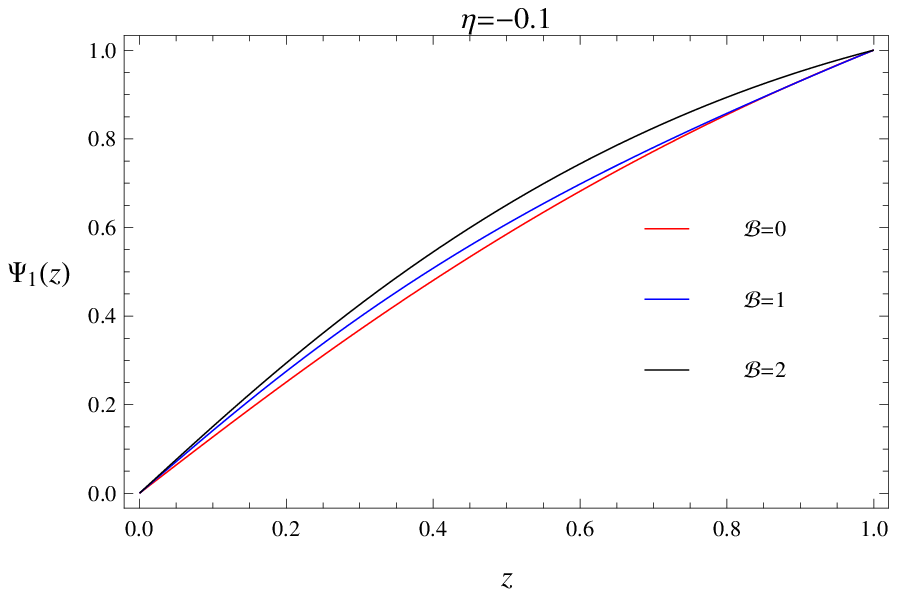}\hspace{0.1cm}
\includegraphics[scale=0.55]{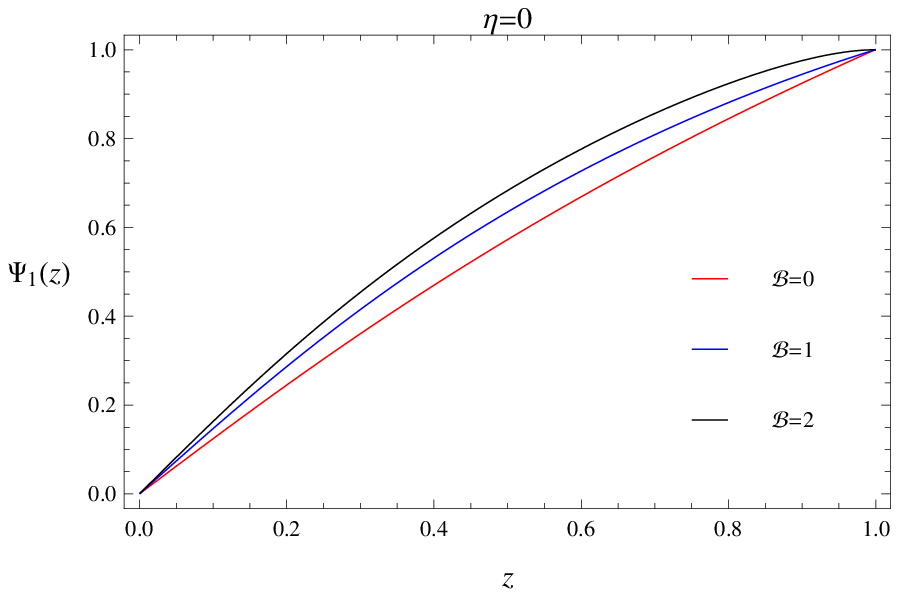}\hspace{0.1cm}
\includegraphics[scale=0.55]{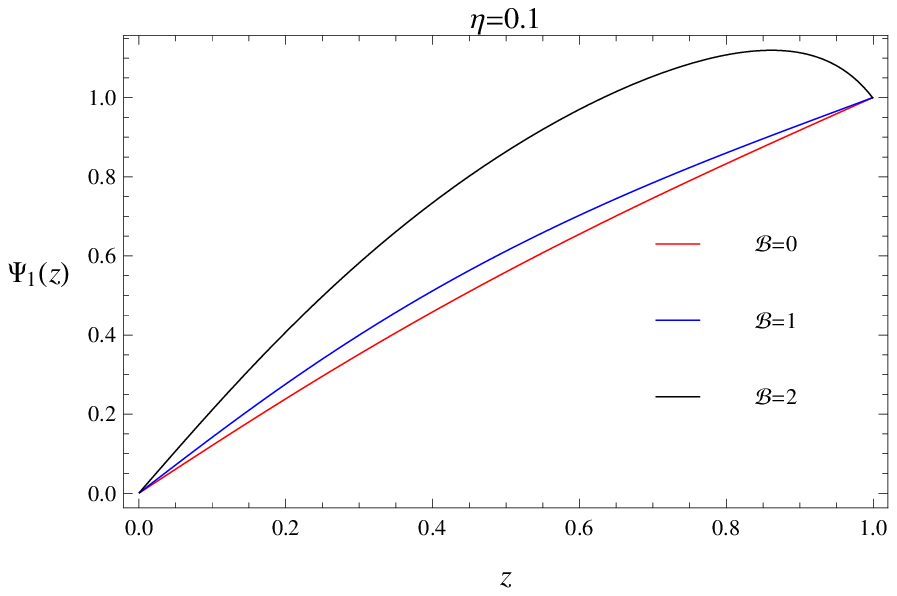}
\caption{\label{fig-zpsi1} The $z$-dependence of the scalar field for various values of the magnetic field $\mathcal{B}$ and coupling constant $\eta$.}}
\end{figure}

The
asymptotic boundary condition   $(z\rightarrow0)$ becomes
\begin{equation}\label{bdyconditon}
\Psi_{1}(z)=\psi_1z+\psi_2z^2~.
\end{equation}
 At the horizon, we have the regularity condition
\begin{equation}\label{initialcondition}
\Psi_{1}'(1)=\frac{2-\mathcal{B}(1+\eta
\mathcal{B}^2)}{(3-\frac{\lambda^2+\mathcal{B}^2}{4})(1-\eta\lambda^2)}\Psi_1(1)~.
\end{equation}
%\begin{equation}\label{initialcondition}
%\Psi_{1}'(1)=\frac{2-\overline{k}^2\mathcal{B}^2(1+\eta  \mathcal{B}^2)}{3(1-\eta \lambda^2)}\Psi_1(1).
%\end{equation}
We will  investigate the  $z-$dependent profile of
the scalar field perturbation, solving equation (\ref{eqPsi1}) focusing on the case of scaling dimension $\Delta =1$
(setting $\psi_2=0$).
%\begin{equation}
%\Psi_{1}'(1)=\frac{2-\mathcal{B}^2(1+\eta  \mathcal{B}^2)}{3(1-\eta \lambda^2)}\Psi_1(1)
%\end{equation}
%\subsection{Numerical Result}
For a numerical solution, we impose the boundary condition $\Psi_1(1)=1$, and solve
(\ref{eqPsi1}) with the use of a shooting method. Notice that the boundary condition at the horizon (\ref{initialcondition}) depends on the coupling constant $\eta$. The
$z-$dependence of the scalar field for different values of the magnetic field is shown
in Fig.~\ref{fig-zpsi1}. We see that as the  magnetic field
becomes stronger, the  corresponding  scalar perturbation in the
bulk becomes bigger.  This behavior  is enhanced by the
increasing strength of the interaction.

\begin{figure}
\center{
\includegraphics[scale=0.75]{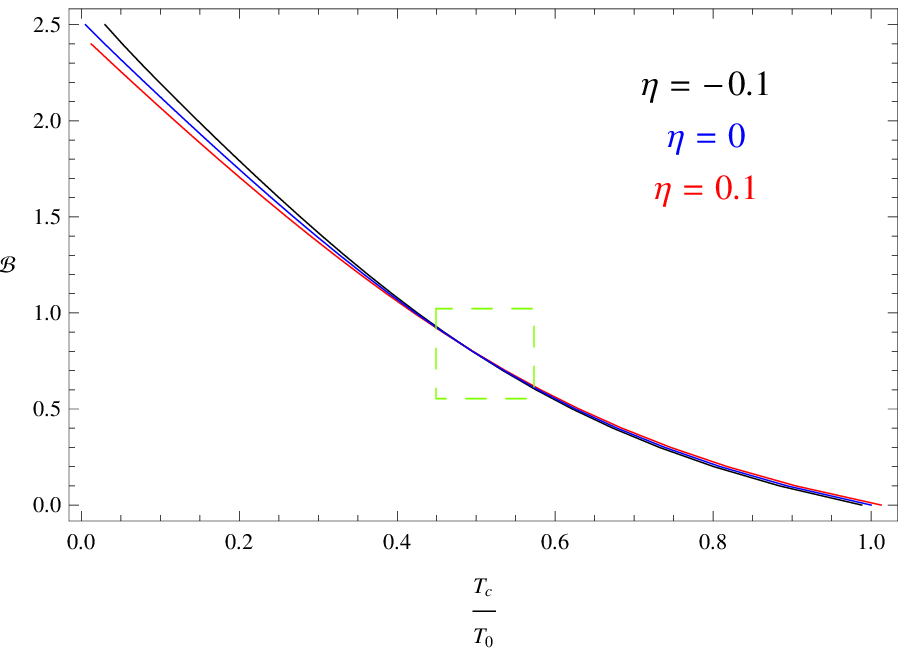}\hspace{1cm}
\includegraphics[scale=0.82]{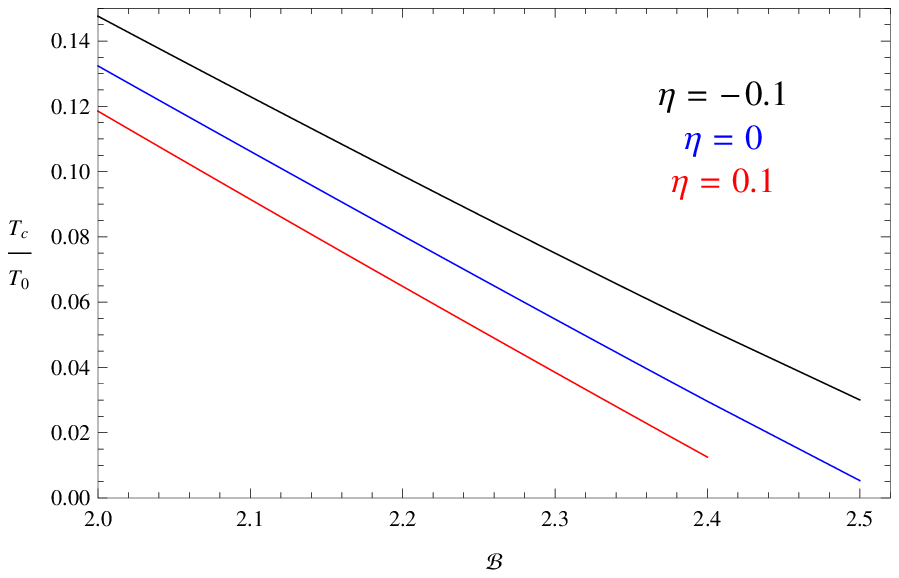}\hspace{1cm}
\caption{\label{fig-TB} The critical temperature as a function of
magnetic field. The low T behavior is enlarged in the right panel. The enlarged view in the green rectangle region is shown in Fig.\ \ref{fig-TBb}.}}
\end{figure}

In Fig.~\ref{fig-TB} we  plot the phase diagram
of the critical magnetic field and critical
temperature. Above the lines, the holographic
system will change into the normal phase.
We use the critical temperature $T_0$ for
$\mathcal{B}=0, \eta=0$ as a scale in the figure.
The phase diagram reveals some very interesting
features.

Stronger magnetic fields correspond to lower
critical temperatures. This behavior is expected,
and for $\eta =0$ it has been discussed in
\cite{Maeda:2009vf,Nakano:2008xc}. However in our
case, as the strength of the higher-derivative coupling
increases, smaller critical temperatures at
strong enough magnetic field
 can be probed as can be seen on the right panel of Fig.~\ref{fig-TB}. To see the behavior of the
system at the quantum critical point ($T=0$), we need to solve the
full back-reacted system of Einstein-Maxwell-scalar field
equations. Note that as the coupling $\eta $ gets larger,
for fixed critical temperature, the critical magnetic field $B_c$,
above which the superconductor enters its normal phase,
decreases.

%One of the reasons of introducing the generalized potential term in condensed matter physics is to account for
%the FFLO states. These states appear in high-field superconductors
%when a strong magnetic field is applied, as Fulde and Ferrell
%\cite{Fulde} and Larkin and Ovchinnikov \cite{Larkin} showed
%independently. Their main
%feature is exhibiting an order parameter which is not a constant,
%but it has a space variation. It was shown that if the
%coefficients in the functional are
%proportional to the magnetic field $B$, and the
%coefficient $\gamma$ of the gradient term in the Landau-Ginzburg
%functional is negative, then the maximum transition temperature
%corresponds to a space dependent state, and is larger than the
%transition temperature of the homogeneous state.
%
%In our case, we recognize some similarities with the
%generalized Landau-Ginzburg theory. Looking at the right panel of
%Fig.~\ref{fig-TB} we see that in the low temperature limit and for
%a strong magnetic field our droplet solutions with negative coupling
%$\eta $ have higher critical temperatures than the solutions with
%positive coupling.

\begin{figure}
\center{
\includegraphics[scale=1]{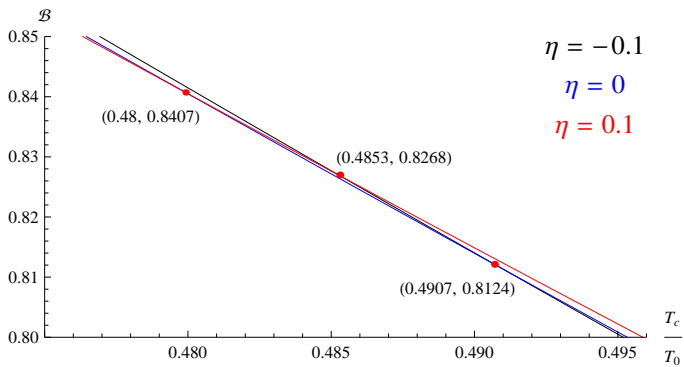}\hspace{1cm}
\caption{\label{fig-TBb} The critical temperature as a function of
magnetic field in the green rectangle region of Fig.\ \ref{fig-TB}.}}
\end{figure}

So far we have concentrated on the strong
magnetic field regime exhibited in
Fig.~\ref{fig-TB}. Let us now look at the left panel of
Fig.~\ref{fig-TB} in the low magnetic field regime.
We observe that, unlike strong magnetic fields, the critical temperature is larger
if the higher-derivative coupling is stronger.
This is consistent with the results obtained below in section \ref{sec:4}.
There is a transition region marked in the inserted
box in the left part of Fig.~\ref{fig-TB}, which
we enlarge in Fig.~\ref{fig-TBb}. We find that
the cross point between lines of $\eta =-0.1$ and
$\eta =0$ occurs at $\mathcal{B}\approx 0.8124$. Below
this value of the magnetic field, the critical
temperature is smaller for negative coupling $\eta$.
Above this value,
the critical temperature for $\eta =-0.1$ becomes
higher than that of the minimal coupling. The
cross point between lines of $\eta =-0.1$ and
$\eta =0.1$ appears at $\mathcal{B}\approx 0.8268$.
Above this transitional strength of the magnetic
field, the critical temperature for $\eta =-0.1$
is higher than the positive coupling. The
transitional strength of the magnetic field
between $\eta =0.1$ and the minimal coupling is at
$\mathcal{B}\approx 0.8407$, above which the
critical temperature for minimal coupling becomes
higher.

One of the reasons of introducing the generalized potential term  was to account for
the FFLO states. These states appear in high-field superconductors
when a strong magnetic field is applied, as Fulde and Ferrell
\cite{Fulde} and Larkin and Ovchinnikov \cite{Larkin} showed
independently. Their main
feature is exhibiting an order parameter which is not a constant,
but has a space variation.
Unfortunately, we cannot
compare the transition temperature of our inhomogeneous solution
to that of a homogeneous solution, because the latter does not
exist in our system.\footnote{It is possible to compare
homogeneous and inhomogeneous solutions in the presence of a
magnetic field, if we introduce two gauge fields, as was discussed
in \cite{Alsup:2012ap,Alsup:2012kr}. The first gauge field couples
to the scalar sourcing a charge condensate below a critical
critical temperature, whereas the second gauge field incorporates
a magnetic field that couples to the spin in the boundary theory.}
We will comment further on this point in section \ref{sec:5}.


In the above calculations, we have considered an exact solution of
the field equations  (dyonic black hole), which includes the
back-reaction of the electromagnetic field. Thus, our results for
the critical temperature are exact in the entire parameter space
depicted in Fig.\ \ref{fig-TB}. As we lower the temperature below
the critical temperature, the field equations become considerably
more involved. We shall concentrate on the case of a weak magnetic
field where the backreaction on the metric can be safely ignored.
In this regime, the critical temperature is finite and there is a
region below $T_c$ which is well above zero temperature so that
the no-backreaction approximation remains valid. In the strong
magnetic field limit, there is no region below $T_c$ where the
backreaction can be ignored, because the critical temperature is
already close to zero.

\section{Below the critical temperature}
\label{sec:4}

%In this work we will not discuss the backreaction of the scalar field to the metric and we will work in the probe limit.
In this section we perform calculations below the critical temperature. We concentrate on the weak magnetic field limit. To simplify the calculations, we also focus on the probe limit by ignoring the back-reaction to the metric.
Thus the metric takes the form of a planar Schwarzschild AdS black hole,
\begin{equation}\label{schwar}
 ds^2= \frac{1}{z^2} \left[ -h(z) dt^2 + \frac{dz^2}{h(z)} + dx^2 +
 dy^2 \right] \ \ , \ \ \ \
h(z)=1 - z^3~.
\end{equation}
We will study the condensation of the scalar
field and calculate the critical temperature both
analytically and numerically in the weak magnetic field limit
$A_i\to 0$ $(i=1,2,3)$. This results in a solution which is approximately spatially
independent. Thus, we consider a spherically symmetric ansatz
\begin{equation}
\Psi=\Psi(z)~,~~~A_\mu=A_t(z)dt~.
\end{equation}
Note that the potential term (\ref{holpot}) in the weak magnetic field limit ($\mathcal{B} \to 0$) only contains gradients in the
radial direction. Therefore, it merely induces interactions between the electric field and the scalar field.

The Maxwell equation (\ref{Maxw}) and the scalar field
equation (\ref{eqPsi}) in the black hole background (\ref{schwar})
reduce to, respectively,
\begin{eqnarray}\label{Maxw2z}
A_t''+\left[\frac{\frac{2\eta}{r_+^2} z^3\Psi A_t^2(z\Psi
h'-2h(\Psi+z\Psi'))}
{-\frac{2\eta}{r_+^2} z^4h\Psi^2A_t^2+z^2h^2(1+2\eta z^2h\Psi'^2)}+\frac{2\eta z^3h\Psi'(z
h'\Psi'+2h(\Psi'+z\Psi''))}{-\frac{2\eta}{r_+^2} z^4\Psi^2A_t^2+z^2h(1+2\eta z^2h\Psi'^2)}\right]A_t'\no\\-
\left[\frac{2\Psi^2(1+\frac{\eta}{r_+^2} z^4A_t'^2)}{-\frac{2\eta}{r_+^2} z^4\Psi^2A_t^2+z^2h(1+2\eta z^2h\Psi'^2)}\right]A_t=0~,
\end{eqnarray}
\begin{eqnarray}\label{eqPsi2z}
\Psi''+\left[\frac{h'}{h}-\frac{2(1+\frac{\eta}{r_+^2} z^4A_t'^2+\frac{\eta}{r_+^2} z^5A_t'A_t'')}{z(1-\frac{\eta}{r_+^2} z^4A_t'^2)}\right]\Psi'
+\left[\frac{A_t^2}{r_+^2h^2}-\frac{m^2}{z^2h(1-\frac{\eta}{r_+^2} z^4A_t'^2)}\right]\Psi=0~,
\end{eqnarray}
where the prime denotes differentiation with respect to $z$. Note
that  for vanishing coupling $\eta $, the above field equations
coincide with the corresponding equations in
\cite{Hartnoll:2008vx}.

As in \cite{Hartnoll:2008vx}, the crucial feature is the last term in
(\ref{eqPsi2z}) which is a direct coupling of the scalar potential
to the scalar field. This term comes from the effective mass term
which reads\footnote{ $m_{\text{eff}}^2$ is the coefficient of $\Psi^2$ in
the action (\ref{action}).}
\begin{eqnarray}\label{meff}
m_{\text{eff}}^2=m^2-\frac{z^2 A_t^2(1-\frac{\eta}{r_+^2} z^4A_t'^2)}{r_+^2h}
\end{eqnarray}
and besides the scalar potential it also depends on the coupling
constant $\eta $. This is the gravitational analog
\cite{Gubser:2008px} of an Abelian $U(1)$ symmetry breaking
outside the horizon of the  black hole
 background (\ref{schwar}) as the effective mass \eqref{meff} drops below the Breitenlohner-Freedman bound $m^2_{\text{BF}} = - \frac{9}{4}$.

\begin{figure}[t]
\center{
\includegraphics[scale=0.62]{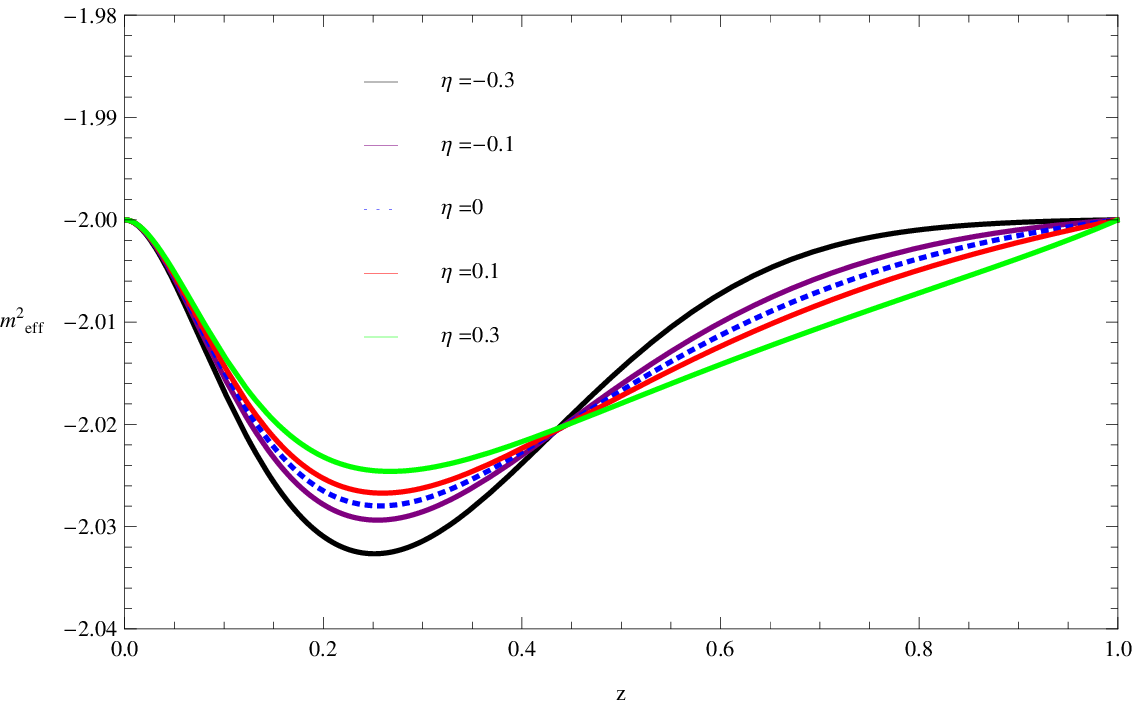}
\includegraphics[scale=0.8]{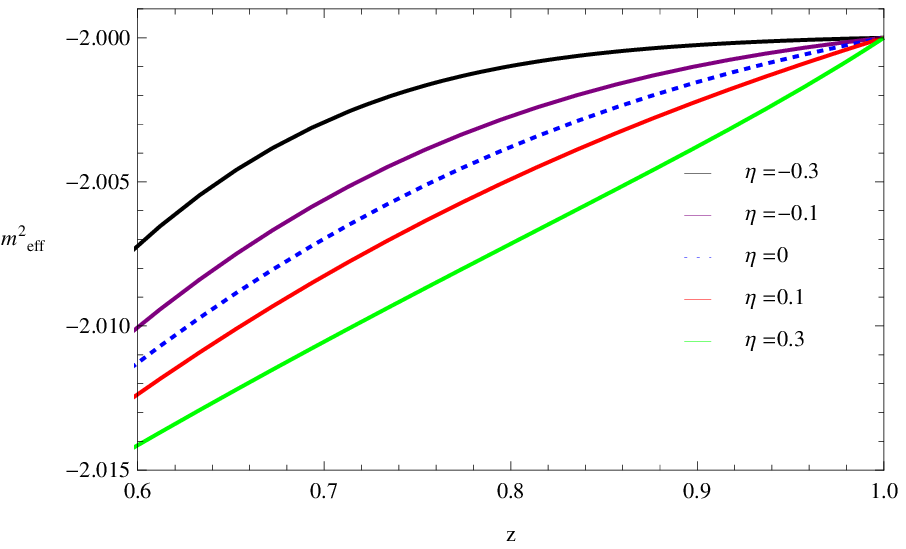}
 \caption{\label{fig-meff1} The value of the
effective mass of the scalar field (\ref{meff}) as a function of $z$ for scaling dimension $\Delta =1$ and various values of the coupling
$\eta $.}}
\end{figure}

\begin{figure}[t]
\center{
\includegraphics[scale=0.8]{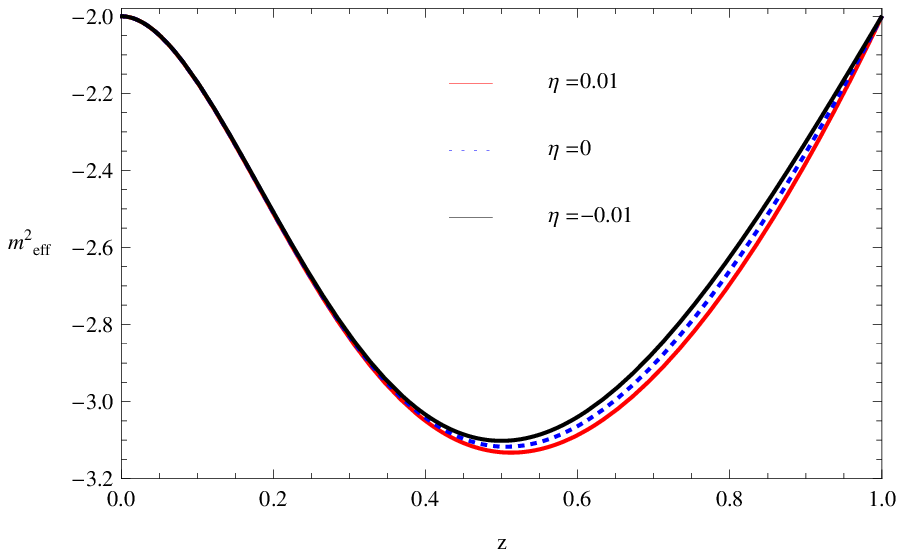}
\includegraphics[scale=0.8]{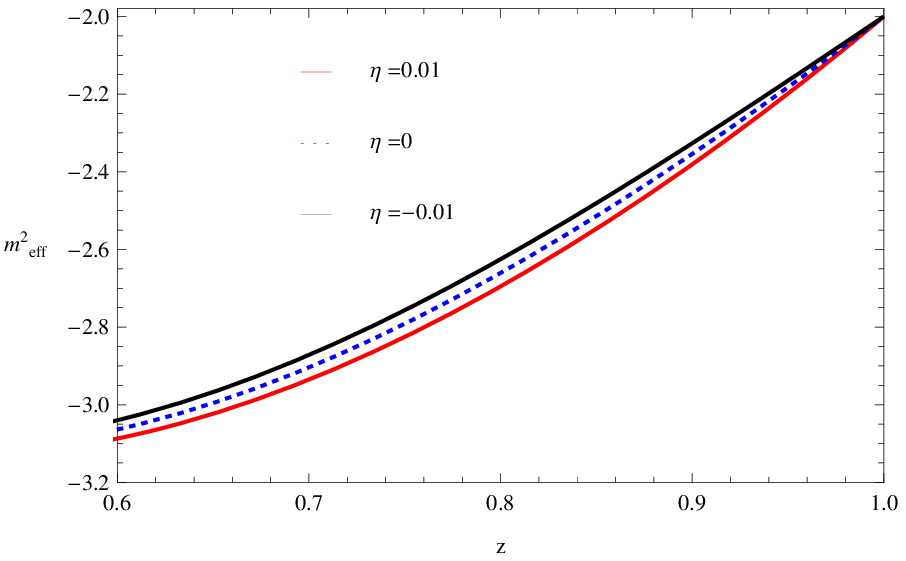}
 \caption{\label{fig-meff2} The value of the
effective mass of the scalar field (\ref{meff}) as a function of $z$ for scaling dimension $\Delta =2$ and various values of the coupling
$\eta $.}}
\end{figure}

 To see the effect of the presence of the
 interaction term, in Fig.~\ref{fig-meff1} we plot the effective
 mass for $m^2=-2$ and scaling dimension $\Delta =1$.
% We have used the value $m^2=-2$ to have a stable black hole background \cite{Breitenlohner:1982jf}.
The left panel
shows that large negative values of the coupling constant $\eta $
make $m_{\text{eff}}^2$ develop a deeper well. However, as we
approach the horizon (right panel) larger values of the coupling
$\eta $ give smaller values of the effective mass, which means that
the condensation is easier to form. Therefore, the effect of the
presence of the interaction term is that as the strength of the
coupling $\eta $ is increased, the system undergoes a phase
transition in a higher critical temperature as can be seen in
Table~\ref{g1-temp1}. The same behavior can be observed in the case of scaling dimension $\Delta =2$ as can be seen in Fig.~\ref{fig-meff2}
and Table~\ref{g1-temp2}.

Next we will calculate the critical temperature analytically and
show that it agrees with our numerical results.

\subsection{Analytic Results}

We will analytically determine the critical temperature for scaling dimension $\Delta =1$ with the use of the method developed in
\cite{Siopsis:2010uq}. Since
the scalar field $\Psi$ vanishes at the critical temperature $T_c$,
the Maxwell equation (\ref{Maxw2z}) reduces to $A_t''=0$, which is solved by
\begin{eqnarray}\label{AtTc}
A_t=\lambda{ r_{+}}(1-z)~,
\end{eqnarray}
where $\lambda r_+^2$ is the charge density. On the other hand, near the boundary,
 we can introduce a function $F(z)$ by defining\footnote{For the other scaling dimension, $\Delta =2$,
 the definition becomes $\Psi=\frac{\langle \mathcal{O}_2 \rangle}{\sqrt{2}r_+}z^2 F(z)$.}
\begin{eqnarray}\label{FTc}
\Psi=\frac{\langle \mathcal{O}_1 \rangle}{\sqrt{2}r_+}z F(z)~,
\end{eqnarray}
where $F(z)$ has been normalized as $F(0)=1$ and $F'(0)=0$ .

Thus, substituting (\ref{AtTc}) and (\ref{FTc}) into  the scalar
field equation (\ref{eqPsi2z}) we obtain
\begin{eqnarray}\label{FeomTc}
F''(z)&-&\left[\frac{5z^3-2}{z(1-z^3)}+\frac{2(1+\eta \lambda^2 z^4)}{z(1-\eta \lambda^2 z^4)}\right]F'(z)
-\left[\frac{3z}{1-z^3}+\frac{2(1+\eta \lambda^2 z^4)}{z^2(1-\eta \lambda^2 z^4)}-\frac{2}{z^2(1-z^3)(1-\eta z^4\lambda^2)}\right]F(z)\nonumber \\
&+&\frac{\lambda^2}{(1+z+z^3)^2}F=0~.
\end{eqnarray}
The above equation can be converted into
\begin{eqnarray}\label{s-leq}
[T(z)F'(z)]'-Q(z)F(z)=0~,
\end{eqnarray}
with
\begin{eqnarray}\label{s-leco}
T(z)&=&(1-z^3)(1-\eta \lambda^2 z^4)~,~~~~~~~Q(z)=z -\lambda^2 \frac{(1-z)(1-\eta \lambda^2 z^4)}{1+z+z^2}-\eta\lambda^2 z^2(5z^3-2)~.
\end{eqnarray}
With the use of Sturm-Liouville theory, the eigenvalue $\lambda$ can be found as the minimum solution of
\begin{eqnarray}\label{lambda1}
\int_0^1
dz\left[T(z)F'(z)^2+Q(z)F(z)^2\right] =0~.
\end{eqnarray}
To proceed, we assume the trial function
\begin{equation}
F(z)\equiv1-\alpha z^2~,
\end{equation}
which satisfies the condition $F(0)=1$ as well as $F'(0)=0$. After integrating over $z$ in (\ref{lambda1}), we find that $\lambda^2$ satisfies the second-order equation
\begin{equation}\label{lambda2}
a_2 \lambda^4 + a_1 \lambda^2 + a_0 = 0~,
\end{equation}
where
\
\begin{eqnarray}\label{lambda2a}
a_2 &=& \eta  \left\{ \alpha^{2}\left[2\sqrt{3}\pi
+{3}\left( -\frac{409}{70}+2\ln3\right)\right]
+2\alpha\left[2\sqrt{3}\pi-3\left( \frac{7}{5}+2\ln3 \right)\right]
+13-12\ln3 \right\}\no\\
a_1 &=& \eta\alpha^{2}\left[ -\frac{110501}{2310}+6\sqrt{3}\pi
+18\ln3\right] + \alpha^2(12\ln3-13) +
\alpha(4\sqrt{3}\pi+12\ln3-36)\no\\
&&+\eta \alpha\left[-\frac{3263}{105}+12\sqrt{3}\pi-36\ln3\right]
+3\eta \left( \frac{97}{7}-12\ln3 \right) +2\sqrt{3}\pi-6\ln3\no\\
a_0 &=& -10\alpha^2+6\alpha-6~.
\end{eqnarray}
For a fixed value of the coupling  $\eta $, we obtain the
minimum value of $\lambda^2$ by varying $\alpha$. Then
the critical temperature is found to be
\begin{eqnarray} \label{critemp}
T_c=\frac{3}{4\pi}r_+=\frac{3}{4\pi\sqrt{\lambda_{\mathrm{min}}}}\sqrt{\rho}~.
\end{eqnarray}
With the use of (\ref{critemp}), the
analytic critical temperatures are summarized in
Table~\ref{g1-temp1}.  In Table~\ref{g1-temp2} we
also summarize the critical temperature for the
case of scaling dimension $\Delta = 2$ calculated
analytically. We see that the critical
temperature increases as the coupling constant $\eta $
increases. This is reasonable, because from the
effective mass (\ref{meff}) and
Fig.~\ref{fig-meff1} (or \ref{fig-meff2}), we learn
that larger $\eta $ corresponds to more negative
effective mass, which in turn implies
that the system becomes more unstable and the
symmetry is easier to be broken.

\subsection{Numerical Results}

The field equations (\ref{Maxw2z}) and
(\ref{eqPsi2z}) were solved numerically with
$m^2=-2$. In Table~\ref{g1-temp1} we list the
critical temperature calculated numerically for
various values of the coupling constant $\eta $,
and compare it with the analytic results
calculated from Eq.\ (\ref{critemp}). Note that for
$\eta =0$ we recover the results of
\cite{Hartnoll:2008vx}. We also list the
corresponding values of $\alpha$. From this
table, we can see that the numerical results and
the analytic results are in good agreement.

\begin{figure}
\center{
\includegraphics[scale=0.78]{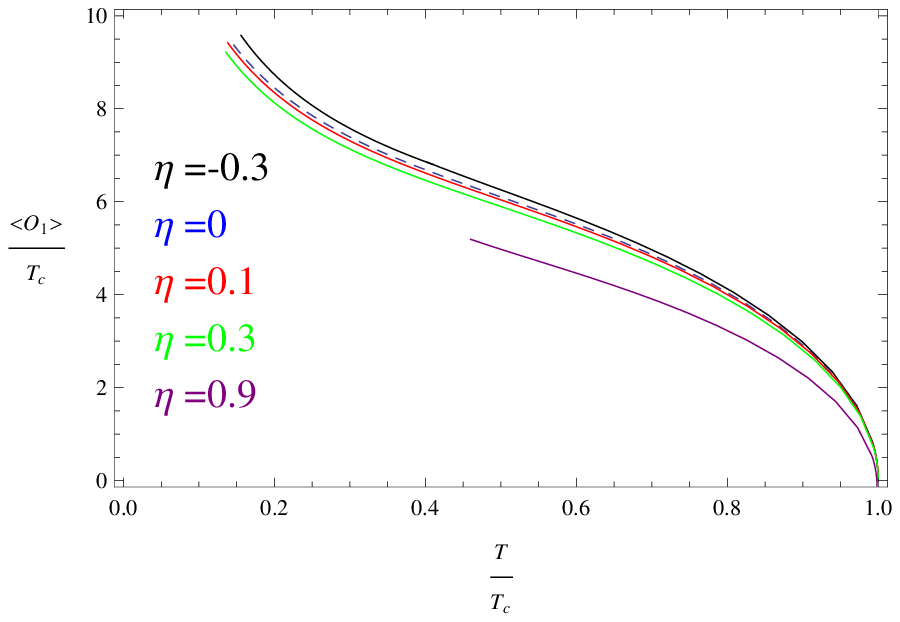}\hspace{0.5cm}
\includegraphics[scale=0.8]{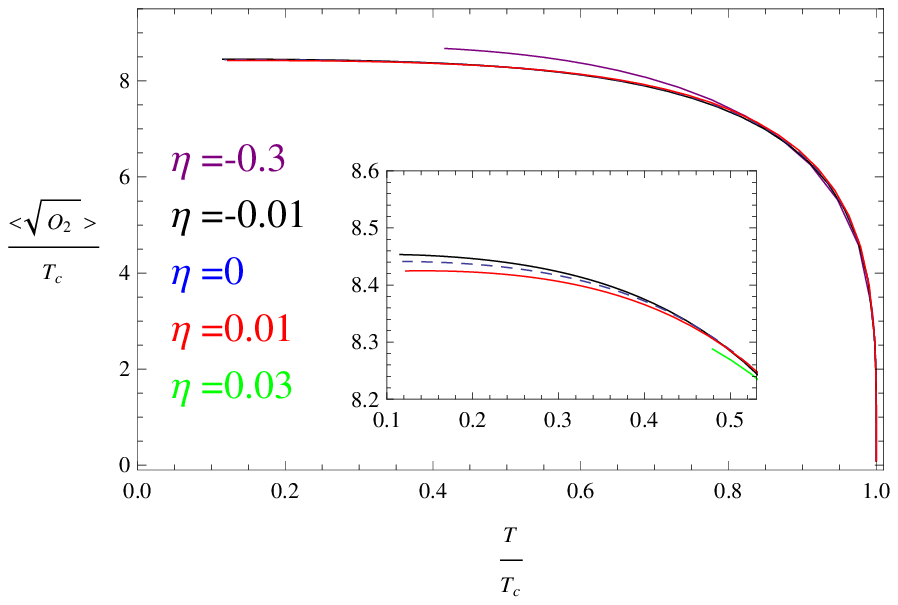}
\caption{\label{fig-conden} The order parameters $\langle\mathcal{O}_1\rangle$ (left panel) and $\langle \mathcal{O}_2\rangle$ (right panel) as functions of
temperature for various values of the coupling $\eta $. }}
\end{figure}

Next we examine the effect of the presence of the
higher-derivative coupling term on the
condensation of the scalar field. In
Fig.~\ref{fig-conden} we show the vacuum
expectation values of the  two operators
$\mathcal{O}_1$ and $\mathcal{O}_2$ versus the
critical temperature. We see that with the
increase of the strength $\eta$ of the
interaction, the gap becomes smaller, which means
that the scalar operator can condense easier when
the coupling $\eta$ is stronger. This agrees with
the property of the critical temperature we
discussed above. For higher $\eta $, our
numerical calculation becomes harder at lower
temperature, however this does not prevent us
from observing the qualitative influence of the
interaction on the condensation. For scaling
dimension $\Delta =1$, this behavior is more
pronounced (see left panel of
Fig.~\ref{fig-conden}). For scaling dimension
$\Delta =2$, we enlarge a local region inserted
into the right panel of Fig.~\ref{fig-conden} to
bring out this behavior. From the critical
temperature behavior obtained with both numerical
and analytic methods, together with the observed
gap behavior, we can conclude that in the weak
magnetic field limit, the greater the strength of
the interaction the easier it is for the
condensation to form. This implies that
in the boundary field theory, which is dual to the
gravitational theory, with stronger
higher-derivative coupling between the $U(1)$
gauge field and the scalar field, the gauge symmetry
can be broken more easily.

\begin{table}
\centering
\begin{tabular}{|c||c||c||c||c||c||c||c||c|}
  \hline
  $\eta $ & -0.3&-0.2& -0.1& 0 & 0.01 & 0.1 & 0.2& 0.3  \\ \hline
  $T_c/\sqrt{\rho}$ (numerical) & 0.2224& 0.2234& 0.2244& 0.2255 & 0.2257 & 0.2267& 0.2280 & 0.2294 \\ \hline
  $T_c/\sqrt{\rho}$ (analytic) & 0.2218& 0.2228& 0.2239& 0.2250 & 0.2251 & 0.2261& 0.2273 & 0.2287 \\
  $(\alpha)$& (0.2868)& (0.2720)& (0.2561)& (0.2389)& (0.2371) & (0.2204)& (0.2006) & (0.1793)\\ \hline
\end{tabular}
\caption{The critical temperature for scaling dimension $\Delta =1$ and various
values of the coupling $\eta $.} \label{g1-temp1}
\end{table}

\begin{table}
\centering
\begin{tabular}{|c||c||c||c||c||c||c||c||c|}
  \hline
 $\eta $ &-0.2& -0.1& 0 & 0.01 & 0.1 & 0.2& 0.3  \\ \hline
  $T_c/\sqrt{\rho}$ (analytic) & 0.1092& 0.1157& 0.1170& 0.1170 & 0.1191& 0.1384 & 0.1506 \\ \hline
\end{tabular}
\caption{The critical temperature for scaling dimension $\Delta =2$ and various
values of the coupling $\eta $.} \label{g1-temp2}
\end{table}

\section{Discussion}
\label{sec:5} We set up a gravitational dual of a holographic
superconductor which included a higher-derivative coupling of
strength $\eta$. Our system had standard composition, consisting
of a  $U(1)$ gauge field, and a complex scalar field coupled to
gravity. The novel feature was a higher-derivative coupling
between the $U(1)$ gauge field and the scalar field. We solved the field
equations and compared the results with the conventional case of a holographic
superconductor ($\eta =0$).
%in the case the scalar field does not backreact on the
%metric. We considered two limits depending on the strength of the
%magnetic field.

In the limit in which the magnetic field is weak, we found a
spatially independent (homogeneous) solution. In the case in which
the coupling strength $\eta $ of the higher-derivative coupling vanishes,
we recovered the results of \cite{Hartnoll:2008vx}. As the
strength of the coupling increases, the gravitational mechanism of
breaking an Abelian $U(1)$ symmetry outside the horizon of a black
hole \cite{Gubser:2008px} becomes more effective, as can be seen
in Fig.~\ref{fig-meff1} and Fig.~\ref{fig-meff2}. Calculating the
critical temperature, both analytically and numerically, we found
that as the strength of the interaction increased, there was an
enhancement of the critical temperature (Tables \ref{g1-temp1} and
\ref{g1-temp2}) at which the scalar condensate formed.

In the presence of a magnetic field, the normal state corresponds to
the background of a dyonic black hole. The field equations possess spatially dependent
(inhomogeneous) solutions. We found that the presence of the higher-derivative
coupling did not affect the $x$-dependent profile of the scalar
field solution. The evolution of the $x$-dependent profile of the scalar field
is given by the solution of equation (\ref{Psi2b}) which depends on the strength of the magnetic field.
As it is discussed in \cite{Albash:2008eh},
when the magnetic field is zero, the $x$-dependence of the condensate
disappears and we recover the homogeneous solution. However, as
the magnetic field is increased, the $x$-dependent profile of the
scalar field acquires a ``thickness" making the condensate more
inhomogeneous.

The presence of the higher-derivative coupling term influences the radial
dependence of the scalar field contributing to the critical
temperature. In Fig.~\ref{fig-TB}, the $(B,T)$ phase
diagram for various values of the coupling  $\eta $ is shown. We can see some
interesting features of our droplet solutions in the low
temperature, strong magnetic field region of the phase diagram
(right panel of Fig.~\ref{fig-TB}). For $\eta >0$ and a fixed value
of the magnetic field we can probe lower temperatures. For $\eta <0$
the condensate forms at a higher critical temperature and at a
stronger magnetic field making the corresponding droplet solution
more inhomogeneous. As we discussed, this
behavior is reminiscent  of the formation of FFLO states, in the case where inhomogeneous states at low
temperature are energetically more favorable.

To have a better understanding of our droplet
solutions in the $T \rightarrow 0$ limit and
their possible connection with the inhomogeneous
FFLO states, we have to improve our numerical
techniques in order to probe larger (positive and
negative) values of the coupling constant $\eta $. We
could also look for solutions with a strong
magnetic field penetrating the whole $xy$-plane
and study the behavior of the system by solving
the full non-linear system of the coupled
Einstein-Maxwell-scalar equations.
Besides the higher-derivative coupling
considered in this work, it is also of interest
to generalize the study on the influence of other
higher-derivative terms in the holographic
superconductor, especially in the presence of the
external magnetic field.

Another interesting problem is to extend this analysis to other
types of holographic superconductors which have more structure,
e.g., the p-wave superconductor. The motivation is two-fold.
Firstly, it would be interesting to see if the effect we found in
the s-wave superconductor, namely that the value of the gap is  reduced as
the strength of the derivative coupling increases, persists in
the case of the richer structure of a  p-wave superconductor.
Secondly, one can study how the presence of the higher derivative terms affects
the behavior of a p-wave superconductor as the strength of the
magnetic field increases and the temperature is lowered. We have
to note however, that even in the s-wave superconductor, the
inclusion of the higher-derivative term brings technical
difficulties so one should expect an increase in the complexity in the case of p-wave
superconductor.

\begin{acknowledgments}
This work is supported partially by the NNSF of
China and the Shanghai Science and Technology
Commission under Grant No.\ 11DZ2260700. G.\ S.\ is
supported in part by the US Department of Energy under Grant No.\
DE-FG05-91ER40627.
\end{acknowledgments}

\end{document}